\documentclass[preprint,aps,prl,superscriptaddress,amssymb]{revtex4}
\usepackage{graphicx,epsf,epsfig}
\usepackage{dcolumn}
\usepackage{bm}
\usepackage{color}
\usepackage{ulem}
\usepackage{epstopdf}

\definecolor{pink}{rgb}{1,0.5,0.5}

\begin{document}
\title{Diluted chirality dependence in edge rough graphene nanoribbon field-effect transistors}
\author{F. Tseng} 
\affiliation{Dept. of Electrical and Computer Engineering,University of Virginia, Charlottesville 22904}
\author{ D. Unluer} \affiliation{Dept. of Electrical and Computer Engineering,University of Virginia, Charlottesville 22904}
 \author{ K. Holcomb}\affiliation{UVa Alliance for Computational Science and Engineering}
\author{ M.R. Stan} \affiliation{Dept. of Electrical and Computer Engineering,University of Virginia, Charlottesville 22904}
 \author{ A.W. Ghosh} \affiliation{Dept. of Electrical and Computer Engineering,University of Virginia, Charlottesville 22904}

\date{\today}
\widetext
\begin{abstract}
We investigate the role of various structural nonidealities on the performance of
armchair-edge graphene nanoribbon field effect transistors (GNRFETs). Our results show that edge roughness dilutes the chirality dependence
often predicted by theory but absent experimentally. Instead, GNRs are classifiable into wide (semi-metallic) vs narrow (semiconducting) strips, defining thereby the building blocks for wide-narrow-wide all-graphene devices and interconnects.
 Small bandgaps limit drain bias at the expense of band-to-band tunneling in GNRFETs.   
We outline the relation between device performance metrics and non-idealities such as width modulation, width dislocations and surface step, and non-ideality parameters such as roughness amplitude and correlation length.

\end{abstract}

\maketitle
Graphene nanoribbons (GNRs) are widely being explored as potential channel materials for nanoelectronic devices ~\cite{rAvou}. GNRs share many advantages with carbon nanotubes (CNTs), including high mobility~\cite{rAvou,rHeer} and minimal top surface dangling bonds. In addition, they raise the possibility of wide-area fabrication using top-down lithographic methods compatible with well established planar silicon CMOS technologies, as well as bottom-up chemical control through edge doping and intercalation. While the structural robustness of CNTs enables high-quality devices, it creates considerable challenges with chemical tunability and placement for circuit-level integration ~\cite{rCass}. Conversely, the structural amorphousness of GNRs opens them up to tunability and control, but increases their sensitivity to structural non-idealities that could degrade
their current voltage (I-V) characteristics.

\begin{figure}[ht!]
\centering
\end{figure}

Crucial to GNRFET analysis is the employment of a well-benchmarked electronic structure theory that circumvents the computational complexity (and bandgap inaccuracies) of LDA-DFT ~\cite{rKie, rKie2}, while capturing the complex chemistry and severe strain effects through transferable parameters that simple tight-binding cannot capture.
Hence in this work we employ a bandstructure model based on EHT to properly capture the distortion chemistry in graphene \cite{rKie}, coupled with a NEGF formalism to explore the role of structural non-idealities on GNRFET performance. Starting with a reconciliation of the anomalous bandgapped structure and absence of prominent chirality dependence in experimental GNRFETs~\cite{rKim,rLi}, we catalogue some of the most influential non-idealities. We also identify critical parameters where these scattering events are most effective, as well as features in the I-Vs that are most robust to these variants. Since our analyses focus on the atomic and electronic structures, we considerably simplify the device electrostatics with a capacitive network, and discuss the complexities of their electrostatic potentials elsewhere \cite{dincer}.

\textit{Edge roughness and absence of chirality dependence.}
Experiments show that chemically derived, ultra-smooth GNRs narrower than ~10nm have `anomalous' bandgaps ~\cite{rLi}, in contrast to straightforward single-orbital tight-binding predictions. Similarly our EHT predictions in Fig.1 shows this suppression of metallicity by the porosity of the GNR edges to transverse waves, amplified for narrow ($<$ 10nm) ribbons by a 3.5$\%$ strain~\cite{rSon} at the edges introduced using geometry-optimized using classical molecular dynamics~\cite{rGau}. Ab-initio~\cite{rSon,rHod} and EHT theories both predict an oscillatory dependence of bandgap on ribbon width, superposed on an inverse power law (Fig.1(c)). Such chirality dependences are absent in experiments of ``ultrasmooth" GNRs ~\cite{rLi} (dashed segments), showing instead a cluster around the 3p+1 nanoribbon results. Our simulations suggest (Fig.2) that this arises from chiral mixing by edge roughness, that promotes the largest transmission band-gap among different width segments.

\begin{figure}[ht]
\centering
\end{figure}

\begin{figure}[ht]
\centering

\end{figure}

Atomistic fluctuations at GNR edges can be classified as width modulation, width dislocation or a combination of both (Fig.2 (a,b)). We model edge roughness with a stochastic distribution following an exponential autovariance function ~\cite{rGhosh} parametrized by a roughness amplitude ($\Delta_m$) and a correlation length ($L_m$). The two classes of edge roughness produce significantly differing results, especially within the first 0.5 eV of the band-edge. Indeed, the modulating width creates significant backscattering at low energies where the mode count is sparser ~\cite{rHeer}. For longer correlations ($L_m = 5$ nm), Figs. 3 demonstrate that {\it{edge roughness is dominated by the largest energy bandgap}} corresponding to the (3p+1) segment. The absence of interband transitions in elastic edge scattering favors the higher HOMO (highest occupied molecular orbital) or lower LUMO (lowest unoccupied molecular orbital) level depending on n or p-type conduction. Faster edge oscillations with a small $L_m = 2$ nm, suppresses higher energy modes, opening an additional 0.50 eV {{pseudo-transmission bandgap}} (Fig.3 (a)). {\it{The pseudo-bandgap explains how the rough-edge band-gaps could exceed the (3p+1) prediction}}, although the effect vanishes for wider segments (Fig. 3b). The net effect is an effective wash-out of chirality dependences, thus classifying AGNRs into {\it{ (a) ultra-wide, semi-metallic, and (b) ultra-narrow, semiconducting segments, in agreement with experiments}}~\cite{rKim,rLi}.

\textit{Device Performance and Design.}
 The salient features in the transmission from even small amounts of edge roughness are evident in the current-voltage characteristics of FETs (Fig.4). I-Vs are calculated using NEGF by integrating the transmission energetically over the operating bias window imposed at the contacts. Since OFF currents
 depend exponentially on threshold voltage while ON currents vary linearly for a ballistic device, the
 former is affected more than the latter. To make a meaningful comparison, however, we adjusted the I-Vs
 to achieve a match of the OFF current and focused instead on the ON-OFF ratio at V$_{ds}$=500mV. 

\begin{figure}[t]
\centering

\end{figure}
As expected, shorter $L_m$ and larger $\Delta_m$s degrade the device mobility, sub-threshold slope, on-current and increases variance in device performance. The transconductance(gm) which scales with mobility can be remedied with wider widths as conduction mode count is denser. Narrower nanoribbons with $L_m > 5$ nm can achieve a faster turn off of drain current with a sub-threshold slopes (S) at 60 mV/dec comparable to that of a corresponding smooth-edge GNR.
While shorter edge roughness correlation lengths degraded sub-threshold slope by at least 20 mV/dec. Similarly OFF current and threshold voltage is influenced as there is a large variance in the transmission  around the transmission band-edges that can potentially open pseudo-transmission bandgaps. The variance in threshold voltage of at least 0.5V is an important design parameter  considered for CMOS circuits such as SRAM (static random access memory)~\cite{rGu}.

As drain voltage has less influence on the GNR channel, threshold voltage is limited by size of the transmission bandgap and any variance due to edge roughness. {\it{An applied drain bias beyond the transmission bandgap creates band-to-band (Zener) tunneling (Fig.4 (b))}}~\cite{rLian}. We note that our channel length (L=10nm) ~\cite{rGuo}is subject to tunneling, which slightly raises the OFF current and lowering ON-OFF ratio. Furthermore performance metrics in the table show that the denser mode count in wider GNRs (2.3nm) are less susceptible to effects of edge roughness as variance in ON-OFF ratio is less compared to the narrower GNR (1.2nm).  
We also found that stiff C-C bonds in graphene lattice prevent out-of-plane step heights even up to 0.5nm from influencing electronic properties and subsequently device performance GNRFETs~\cite{rGuis}. 

In summary, chiral signatures enjoyed by CNTs are washed away for GNRs by edge roughness, especially for narrow ribbons. Hence ribbon width becomes the dominant parameter determining metallicity. This implies that absolute control of GNR widths is unnecessary, and the concept of an all-graphene wide-narrow-wide GNRFET is plausible even with structural nonidealities.
The resulting large band-gaps make them gateable, but open them up to larger device-to-device
fluctuations. Nonetheless, the devices are characterized by excellent sub-threshold slope, Ion/Ioff, and on-current, not to mention the considerable benefits that 2-D electrostatics, Ohmic contacts and high-k dielectrics in a wide-narrow-wide geometry can bring in addition \cite{dincer}.

\textit{Acknowledgments.}
This work was supported in part by a MARCO IFC grant, a UVa FEST award, and an NSF CAREER award.

\pagebreak
Fig 1 caption:
EHT captures the proper GNR chemistry, including (a) mid-gap states near the Fermi energy ($-4.5$ eV) arising from armchair edge dangling bonds (inset: local density of gap states). (b) H-passivation removes edge states, while soft edge boundaries prevent metallicity. A $3.5 \%$ edge strain further enhances the band-gap $E_g$. (c) The experimental lack of chirality and a strong clustering instead around AGNR (3p+1,0) ~\cite{rLi} are attributed to chiral `mixing' by edge roughness that favors the largest transmission bandgap among different segment widths (Fig. 2).

Fig 2 caption:
Edge roughness can be categorized as (a) mixed width ($\Delta_m$=.246nm) vs (b) width dislocations ($\Delta_m$=.123nm). (c) The transmissions with both kinds of roughness ($\Delta_m = .246$ and $.123$ nm, $L_m = 9$ nm, ribbon length L = 18nm) increase at energies corresponding to GNR (3p+1,0) band-edges showing the dominance of the largest band-gapped segment. However this bandgap dominance comes at the expense of suppressed transmission from mode-misalignment and electron backscattering near band-edges .

Fig 3 caption:
Using the same roughness  patterns ($\Delta_m$=.123nm and L$_m$ =2 nm, Lm= 5nm) with multiple seeds on a narrow and wide AGNR we found that(a) narrower ($\sim 1.2$ nm) nanoribbons (GNR(10,0)) with more roughness (L$_m$ 2nm) suppresses transmission near band-edges to create psuedo-bandgap in the transmission.  (b) While on a wider($\sim 2$ nm) nanoribbon (GNR(19,0) with L$_m$= 2nm raises the transmission at the dominant 3p+1 GNR band-edge.

Fig 4 caption:
(a)Variance in drain current versus gate voltage characteristics for a narrower GNR(10,0) under multiple seeded roughness parameters ( $\Delta_m$=.123nm and L$_m$ =2 nm (shaded) , L$_m$= 5nm (hatched)) with average OFF currents pinned at V$_G$=0. Better quality GNR edges (L$_m$=5nm) have less ON current and transconductance variance . (b) Threshold voltage is constrained by any variance in the transmission bandgap before additional applied V$_{ds}$ causes Zener tunneling. Bottom table shows certain device performance metrics for wide and narrow GNR channel FETs.

\pagebreak
\begin{figure}[ht!]
\centering
{\epsfxsize=5in\epsfbox{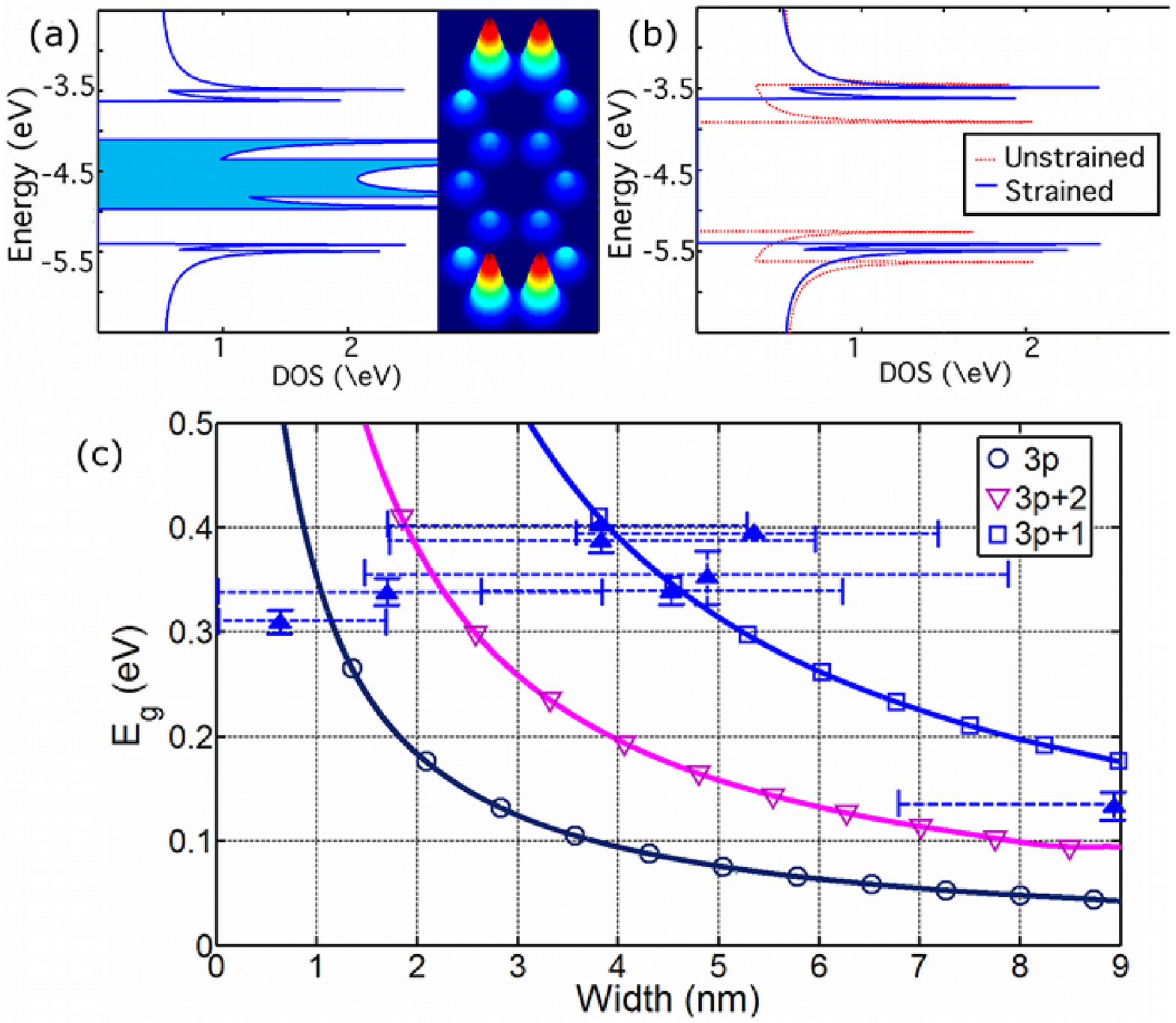}} 

\end{figure}

\begin{figure}[ht!]
\centering
{\epsfxsize=5in\epsfbox{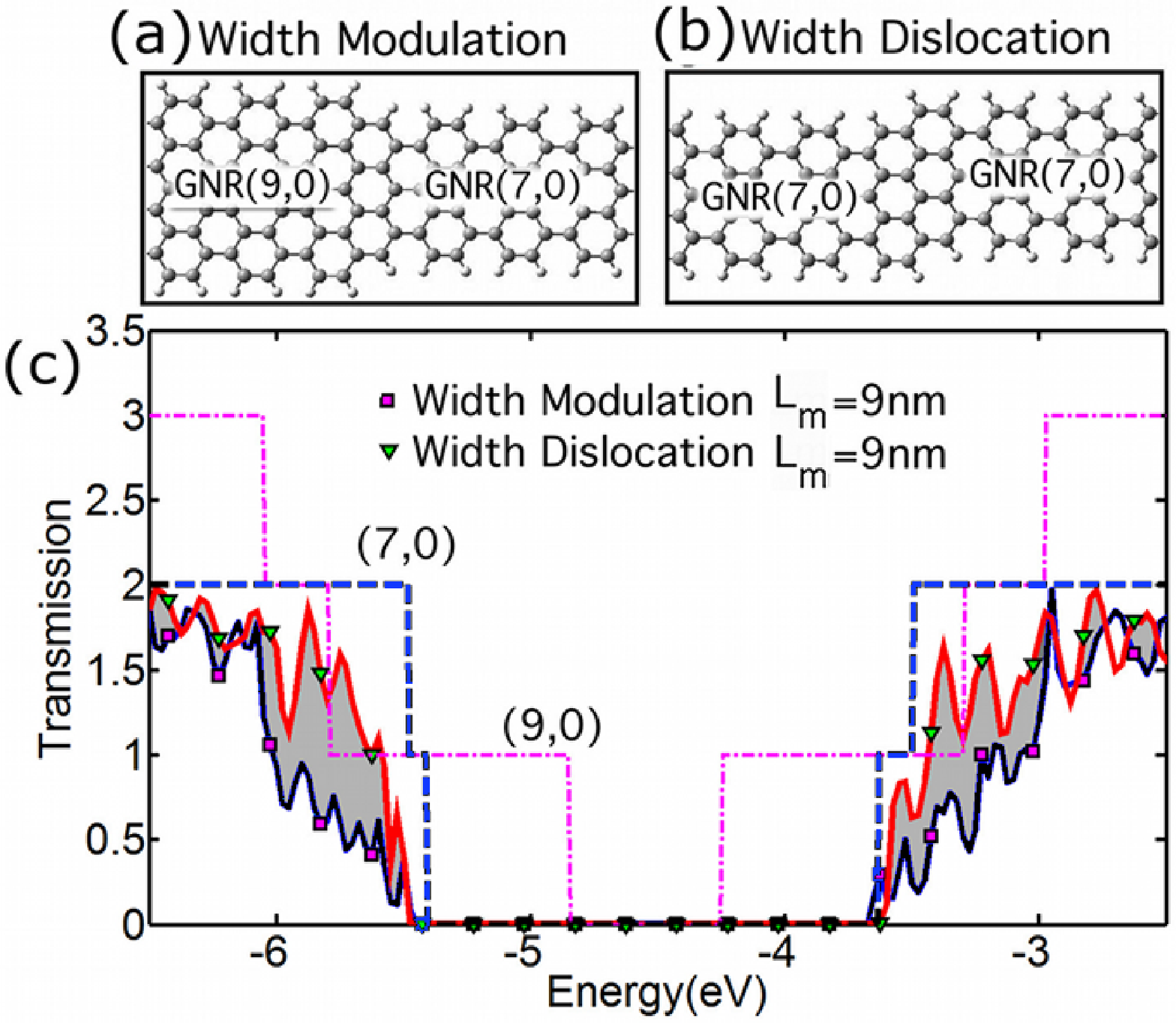}} 

\end{figure}

\begin{figure}[ht!]
\centering
{\epsfxsize=5in\epsfbox{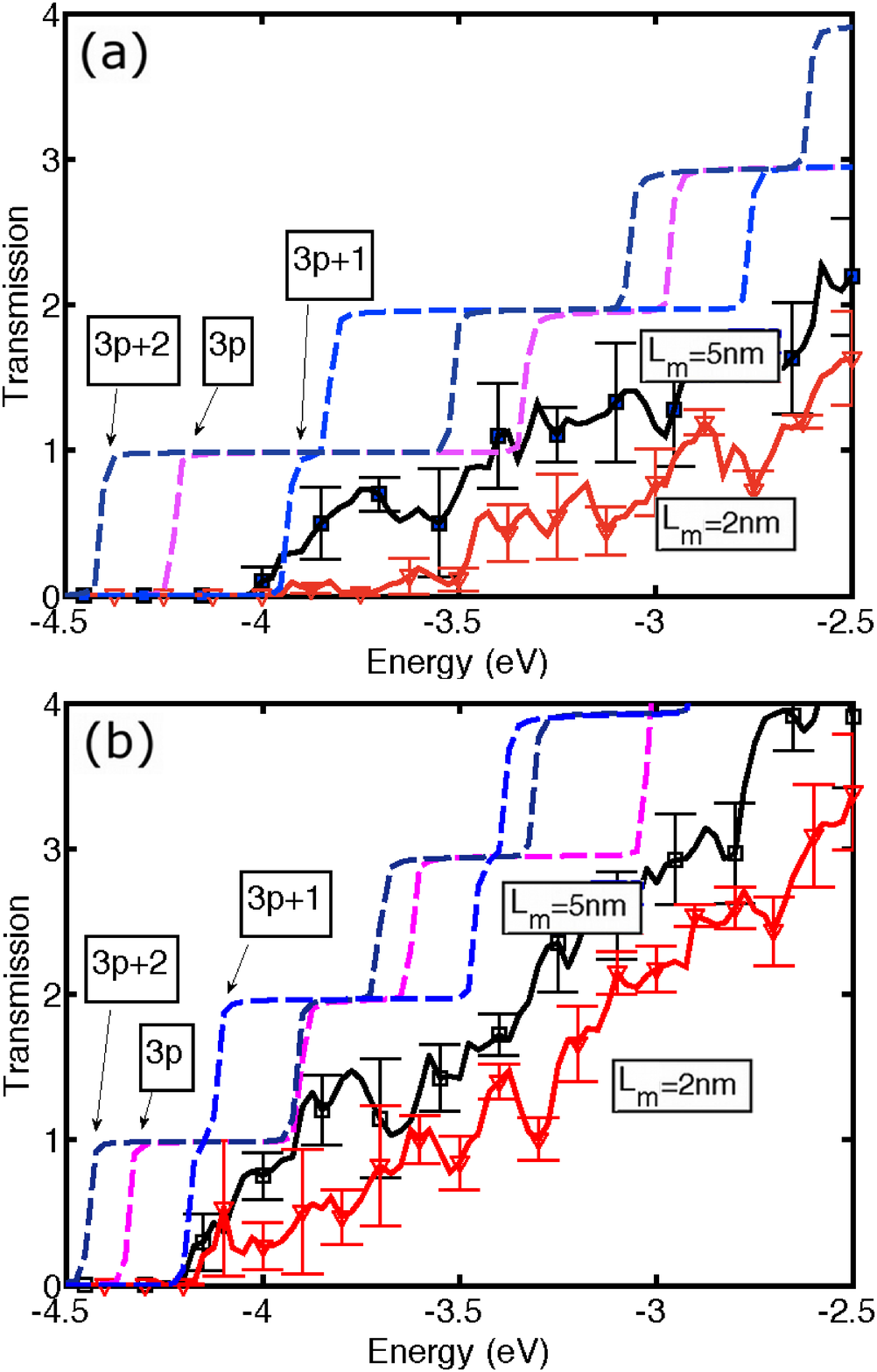}} 

\end{figure}

\begin{figure}[ht!]
\centering
{\epsfxsize=5in\epsfbox{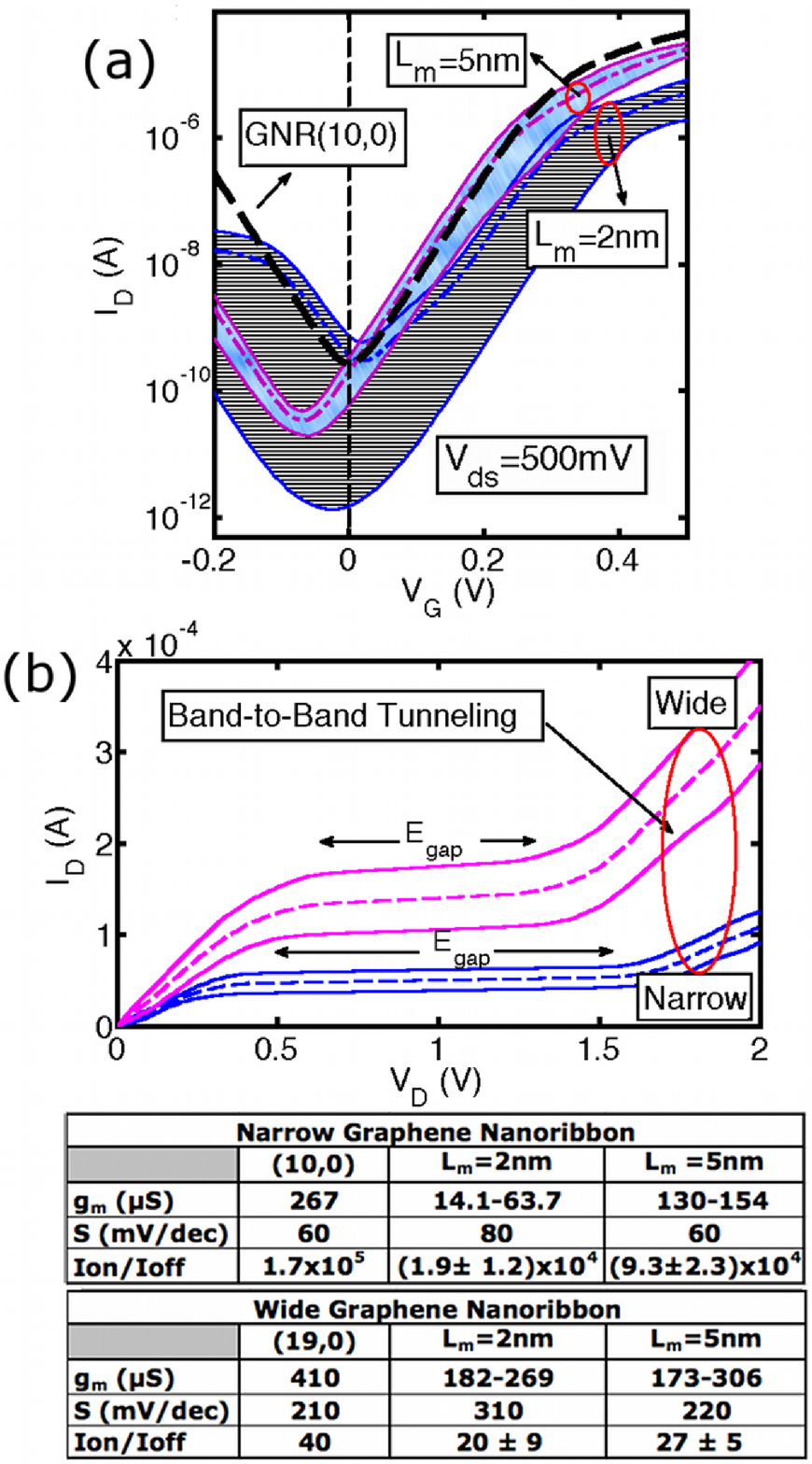}} 

\end{figure}

\end{document}